\documentclass[prx,aps,superscriptaddress,twocolumn,longbibliography]{revtex4-1}
\usepackage[utf8]{inputenc}
\usepackage{amsmath}
\usepackage{natbib}
\usepackage{graphicx}
\usepackage{color}

\begin{document}

\title{Topologically Enhanced Harmonic Generation in a Nonlinear Transmission Line Metamaterial}

\date{\today}

\author{You Wang}

\affiliation{Division of Physics and Applied Physics, School of Physical and Mathematical Sciences,\\
Nanyang Technological University, Singapore 637371, Singapore}

\author{Li-Jun Lang}

\affiliation{Division of Physics and Applied Physics, School of Physical and Mathematical Sciences,\\
Nanyang Technological University, Singapore 637371, Singapore}

\author{Ching Hua Lee}

\affiliation{Institute of High Performance Computing, A*STAR, Singapore 138632, 
Singapore}

\affiliation{Department of Physics, National University of Singapore, Singapore 
117551, Singapore}

\author{Baile~Zhang}
\email{blzhang@ntu.edu.sg}

\affiliation{Division of Physics and Applied Physics, School of Physical and Mathematical Sciences,\\
Nanyang Technological University, Singapore 637371, Singapore}

\affiliation{Centre for Disruptive Photonic Technologies, Nanyang Technological University, Singapore 637371, Singapore}

\author{Y.~D.~Chong}
\email{yidong@ntu.edu.sg}

\affiliation{Division of Physics and Applied Physics, School of Physical and Mathematical Sciences,\\
Nanyang Technological University, Singapore 637371, Singapore}

\affiliation{Centre for Disruptive Photonic Technologies, Nanyang Technological University, Singapore 637371, Singapore}

\begin{abstract}
  Nonlinear transmission lines (NLTLs) are nonlinear electronic circuits used for parametric amplification and pulse generation.  It has previously been shown that harmonic generation can be enhanced, and shock waves suppressed, in ``left-handed'' NLTLs, a manifestation of the unique properties of left-handed media.  We show experimentally that harmonic generation in a left-handed NLTL can be greatly increased by the presence of a topological edge state, using a nonlinear circuit analogue of the Su-Schrieffer-Heeger (SSH) lattice.  Recent studies of nonlinear SSH circuits have investigated the solitonic and self-focusing behaviors of modes at the fundamental harmonic.  We find that frequency-mixing processes in an SSH NLTL have important effects that were previously neglected.  The presence of a topological edge mode at the first harmonic can produce strong higher-harmonic signals that propagate into the lattice, acting as an effectively nonlocal cross-phase nonlinearity.  We observe maximum third-harmonic signal intensities five times that of a comparable left-handed NLTL of a conventional design, and a 250-fold intensity contrast between the topologically nontrivial and trivial lattice configurations. This work may have applications for compact electronic frequency generators, as well as advancing our fundamental understanding of the effects of nonlinearities on topological states.
\end{abstract}

\maketitle

Topological edge states---robust bound states guaranteed to exist at the boundary between media with ``topologically incompatible'' bandstructures---were first discovered in condensed matter physics \cite{Bernevig2013}.  Recently, electronic LC circuits have emerged as a highly promising method of realizing these remarkable phenomena \cite{Ningyuan2015, Albert2015, lee2017, imhof2017, goren2018, zhu2018, hadad2018, serragarcia2018}.  Compared to other classical platforms like photonics \cite{Marin2009Nature,tpreview2014,tpreview2017,tpreview2018}, acoustics \cite{yang2015,fleury2016,He2016}, and mechanical lattices \cite{R1Q3_1, tmreview2016,lee2018}, which have also been used to realize topologically nontrivial bandstructures and topological edge states, electronic circuits have several compelling advantages: extreme ease of experimental analysis; the ability to fabricate complicated structures via printed circuit board (PCB) technology; and the intriguing prospect of introducing nonlinear and/or amplifying circuit elements to easily study how topological edge states behave in novel physical regimes.  Notably, circuits have been used to study the Su-Schrieffer-Heeger (SSH) chain (the simplest one-dimensional topologically-nontrivial lattice) \cite{ssh1979,lee2017}, nonlinear SSH chains supporting solitonic edge states \cite{hadad2018}, two-dimensional topological insulator lattices \cite{Ningyuan2015}, and the corner states of high-order topological insulators \cite{imhof2017,serragarcia2018}.

One of the most interesting questions raised by the emergence of topologically nontrivial classical lattices is how topological edge states interact with nonlinear media.  Previous studies have focused on nonlinearity-induced local self-interactions in the fundamental harmonic, which can give rise to solitons with anomalous plateau-like decay profiles in nonlinear SSH chains \cite{hadad2016,hadad2018}, or chiral solitons in two-dimensional lattices \cite{Lumer2013,Ablowitz2014, Leykam2016, Lumer2016, Kartashov2016, Gulevich2017}.  It has also been suggested that topological edge states in nonlinear lattices could be used for robust traveling-wave parametric amplification \cite{Peano2016}, optical isolation \cite{Zhou2017}, and other applications \cite{Rosenthal2018,chacon2018,kivshar2018a,kivshar2018b}.

\begin{figure*}
  \includegraphics[width=0.99\textwidth]{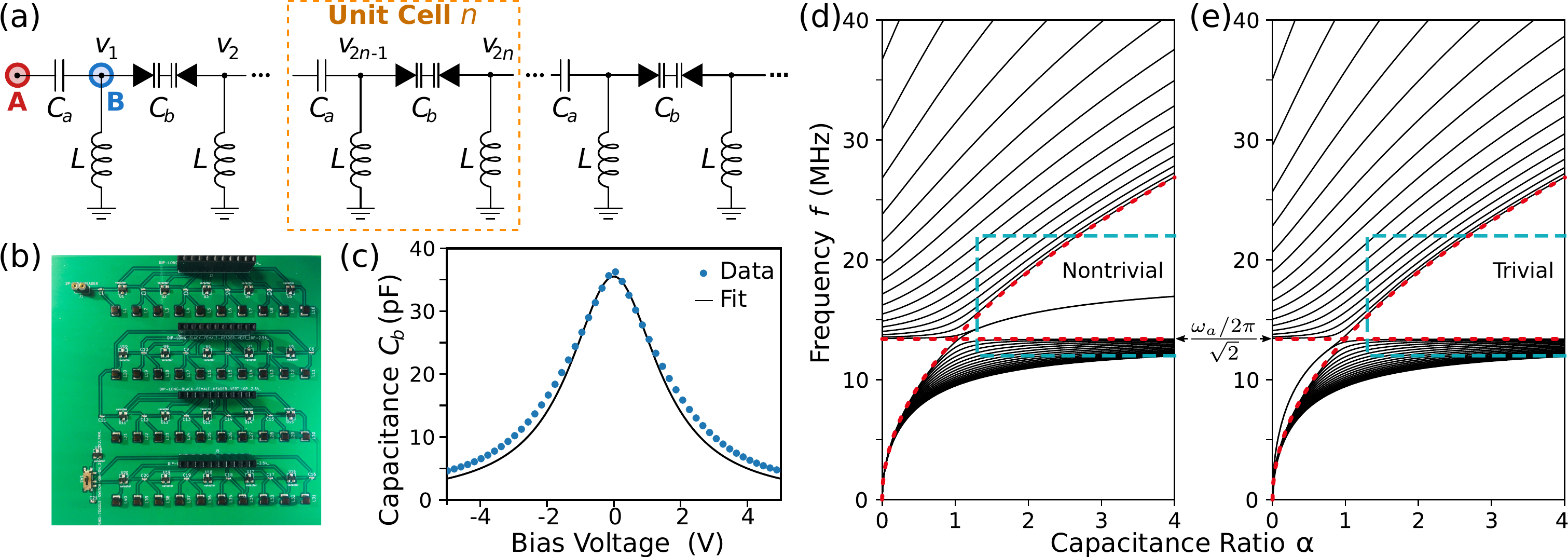}
  \caption{(a) Schematic of a left-handed transmission line circuit with alternating capacitances: linear capacitors with capacitance $C_a$, and back-to-back varactors with nonlinear capacitance $C_b$.  The capacitances act like hoppings in a nonlinear Su-Schrieffer-Heeger (SSH) model.  An input voltage is applied at points $A$ or $B$ to probe the topologically trivial or nontrivial lattice. (b) Photograph of the printed circuit board.  (c) Capacitance $C_b$ versus bias voltage.  Dots are calculated from varactor manufacturer data, and the solid curve is the fit based on Eq.~\eqref{nonlin}.  (d)--(e)  Calculated eigenfrequencies of a finite closed linear circuit, versus the capacitance ratio $\alpha = C_a/C_b$.  The characteristic frequency is $f_a = \omega_a/2\pi \approx 19\,\mathrm{MHz}$, and the lattice has 40 sites.  Two cases are shown: (d) $C_a$-type capacitors at the edge, for which the $\alpha > 1$ gap is topologically nontrivial; (e) $C_b$-type capacitors at the edge, for which the $\alpha > 1$ gap is trivial.  The red dotted curves indicate the band-edge frequencies $f_a/\sqrt{2}$ and $\sqrt{\alpha/2}\,f_a$.  The blue dashes indicate the operating regime of the nonlinear circuit, with $\alpha \approx 1.3$ in the linear (low-voltage) limit and $\alpha$ effectively increasing with voltage amplitude.}
  \label{fig:Circuit}
\end{figure*}

In this paper, we report on the implementation of a nonlinear SSH chain based on a left-handed nonlinear transmission line (NLTL) \cite{Cullen1958,Tien1958,Landauer1960IBM,Landauer1960AIP,Lai2004, Kozyrev2005, Kozyrev2008, Powell2009}, in which the topological edge state induces highly efficient harmonic generation.  The first harmonic mode is localized to the lattice edge, similar to a linear topological edge state, whereas the higher-harmonic waves propagate into the lattice bulk, with voltage amplitudes reaching over an order of magnitude larger than the first harmonic signal.  The intensity of the generated third harmonic signal has a maximum of $\approx 2.5$ times that of the input first-harmonic signal, compared to $< 0.5$ for a comparable conventional left-handed NLTL without a topological edge state.  The important role played by the topological edge state is further demonstrated by the fact that the third-harmonic intensity is 250 times larger than in a ``trivial'' circuit, which has equivalent parameters but lacks a topological edge state in the linear limit, using the same input parameters.

Although previous studies have emphasized the role of local self-interactions, including in a previous demonstration of a nonlinear SSH circuit based on weakly-coupled LC resonators \cite{hadad2018}, an important feature of our circuit is the decisive role that the higher-harmonic signals play in modulating the first-harmonic modes. The strong intensity of higher-harmonic modes effectively drives the entire lattice, not just the edge, deeper into the nontrivial regime at the first harmonic frequencies.  This is aided by the fact that the left-handed NLTL has an unbounded dispersion curve supporting traveling-wave higher-harmonic modes \cite{Vesalago1968,Lai2004, Kozyrev2005, Kozyrev2008, Powell2009}.

\textit{Circuit design}---The transmission line circuit is shown schematically in Fig.~\ref{fig:Circuit}(a).  It contains inductors of inductance $L$ and capacitors of alternating (dimerized) capacitances $C_a$ and $C_b$.  We will shortly treat the case where the $C_b$ capacitors are nonlinear (the $L$ and $C_a$ elements are always linear).  First, consider the linear limit where $C_b$ is a constant.  We define the characteristic angular frequency $\omega_a=(LC_a)^{-1/2}$, and the capacitance ratio
\begin{equation}
  \alpha=C_a/C_b.
\end{equation}
The case of $\alpha = 1$ corresponds to a standard (non-dimerized) left-handed transmission line.  This type of transmission line is characterized by having sites separated by capactitors, and connected to ground by inductors, rather than vice versa.  Left-handed NLTLs have been shown to be useful for parametric amplification and pulse generation \cite{Lai2004, Kozyrev2005, Kozyrev2008, Powell2009}.

Let us treat the points adjacent to the capacitors as lattice sites, indexed by an integer $k$, and close the circuit by grounding the edges [the left edge is the site labelled A in Fig.~\ref{fig:Circuit}(a)].  Using Kirchhoff's laws, we can show that a mode with angular frequency $\omega$ satisfies \cite{SM}
\begin{equation}
  \left(\mathcal{H} - \frac{1}{\alpha}\right) 
  \begin{pmatrix}
    v_1\\v_2\\v_3\\\vdots
  \end{pmatrix}
  = \left(1 -\frac{\omega_a^2}{\omega^2}\right)
  \begin{pmatrix}
    v_1\\v_2\\v_3\\\vdots
  \end{pmatrix},
  \label{lineqn}
\end{equation}
where $v_k$ denotes the complex voltage on site $k$.  The matrix $\mathcal{H}$ has the form of the SSH Hamiltonian:
\begin{equation}
  \mathcal{H} =
  \begin{pmatrix}
    0&\frac{1}{\alpha}&&&&\\
    \frac{1}{\alpha}&0&1&&&\\
    &1&0&\frac{1}{\alpha}&&\\
    &&\frac{1}{\alpha}&0&\ddots&\\
    &&&\ddots&\ddots&\\
  \end{pmatrix}.
\end{equation}
Thus, the eigenfrequency modes of the circuit have a one-to-one correspondence with the SSH eigenstates.

The band diagram for the linear closed circuit is shown in Fig.~\ref{fig:Circuit}(d).  The lack of an upper cutoff frequency is a characteristic of left-handed transmission lines \cite{Kozyrev2008}.  There is a bandgap in the range $\omega_a/\sqrt{2} < \omega < \sqrt{\alpha/2}\, \omega_a$.  For $\alpha > 1$, the bandgap contains edge states, which are zero-eigenvalue eigenstates of $\mathcal{H}$ that can be characterized via a topological invariant derived from the Zak phase \cite{Bernevig2013}.  The edge state's angular frequency is
\begin{equation}
  \omega_{\mathrm{es}} = \sqrt{\alpha/(1+\alpha)}\, \omega_a.
\end{equation}
Note that the edge state are not at zero frequency, nor do they lie at precisely the middle of the bandgap; this is due to the aforementioned mapping from the circuit equations to the SSH model---specifically, the fact that $\omega$ is not the eigenvalue in Eq.~\eqref{lineqn}.

For $\alpha < 1$, there is a finite bandgap below $\omega_a/\sqrt{2}$, which is topologically trivial and contains no edge states.  If we swap the two types of capacitors, so that the $C_b$-type capacitors are the ones at the edge, then the $\alpha>1$ bandgap is trivial and the $\alpha<1$ bandgap nontrivial, as shown in Fig.~\ref{fig:Circuit}(e).


Next, consider a nonlinear circuit with each $C_b$ capacitor consisting of a pair of back-to-back varactors.  The nonlinear capacitance $C_b$ decreases with the magnitude of the bias voltage (the voltage between the end-points of the capacitor), as shown in Fig.~\ref{fig:Circuit}(c).  For theoretical analyses, it is convenient to model this nonlinearity by
\begin{equation}
  \alpha_{\mathrm{nl}}(t) \approx A + B \, \left[\Delta V(t)\right]^2,
  \label{nonlin}
\end{equation}
where $\alpha_{\mathrm{nl}}(t) \equiv C_a / C_{b}(t)$, and $\Delta V(t)$ is the bias voltage.  The key feature of the nonlinearity is that at higher voltages, the effective value of $\alpha$ increases.  Depending on the chosen boundary conditions, this drives the circuit deeper into the topologically trivial or nontrivial regime.


\textit{Experimental results}---The implemented nonlinear transmission line, shown Fig.~\ref{fig:Circuit}(b), contains a total of 40 sites, or 20 unit cells.  The linear circuit elements have $L = 1.5\,\mu\textrm{H}$ and $C_a = 47\,\textrm{pF}$, so that $\omega_a /2\pi \approx 19\,\mathrm{MHz}$.  By fitting Eq.~\eqref{nonlin} to manufacturer data for the varactors at low bias voltages \cite{SM}, we obtain $A = 1.32$ and $B = 0.51\,\mathrm{V}^{-2}$ (thus, in the linear limit, $\alpha \approx 1.3 > 1$).  The fitted capacitance-voltage relation is shown in Fig.~\ref{fig:Circuit}(c).

We supply a continuous-wave sinusoidal input voltage signal, with tunable frequency $f_{\mathrm{in}}$ and amplitude $V_{\mathrm{in}}$, to either of the points labelled $A$ and $B$ in Fig.~\ref{fig:Circuit}(a). This allows us to study the cases corresponding to Fig.~\ref{fig:Circuit}(d) and (e), which we refer to as the ``nontrivial'' and ``trivial'' lattices respectively (see Methods).  In both cases, the input site is denoted as $k = 0$.

\begin{figure}
  \includegraphics[width=0.49\textwidth]{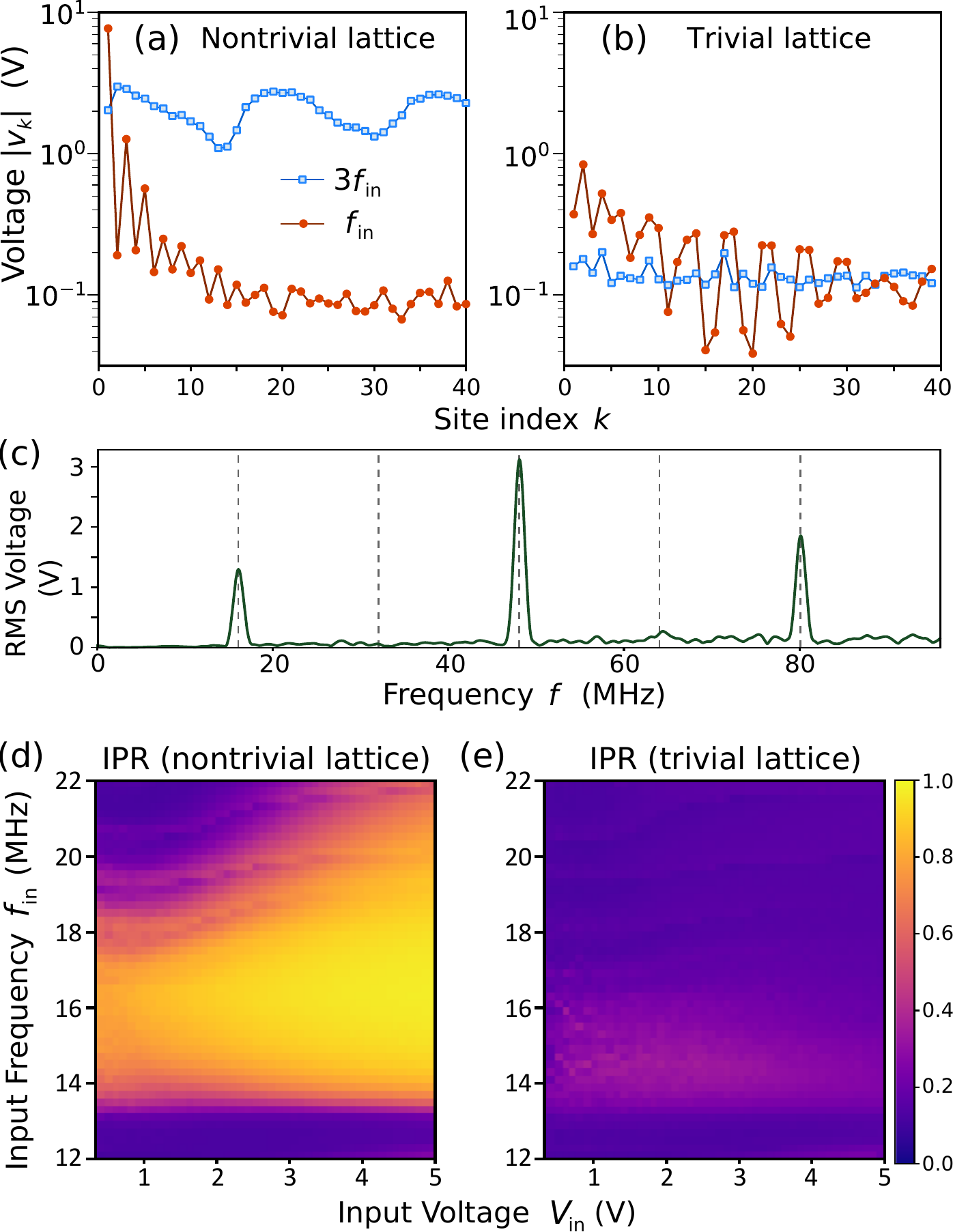}
  \caption{(a)--(b) Magnitude of the first- and third-harmonic voltage signals measured at different lattice sites, for (a) the nontrivial lattice, which has an SSH-like edge state in the linear limit, and (b) the trivial lattice, which has no edge state in the linear limit.  The sinusoidal input signal, applied at the lattice edge (site 0), has frequency $f_{\mathrm{in}}=16\,\mathrm{MHz}$ and amplitude $V_{\mathrm{in}}=2.5\,\mathrm{V}$.  (c) Measured spectrum at site 3 for the nontrivial lattice corresponding to (a).  (d)--(e) Plot of the inverse participation ratio (IPR) versus input frequency $f_{\mathrm{in}}$ and input voltage amplitude $V_{\mathrm{in}}$, calculated from experimental measurements of the first-harmonic signal in the (d) nontrivial and (e) trivial lattices.  Here, $f_{\mathrm{in}}$ is measured in steps of $0.2\,\mathrm{MHz}$, and $V_{\mathrm{in}}$ in steps of $0.1\,\mathrm{V}$.}
  \label{fig:profile}
\end{figure}

A typical set of measurement results is shown in Fig.~\ref{fig:profile}(a)--(c), for $f_{\mathrm{in}}=16\,\mathrm{MHz}$ and $V_{\mathrm{in}}=2.5\,\mathrm{V}$.  On each site $k$, the spectrum of the voltage signal is shown in Fig.~\ref{fig:profile}(c), with prominent peaks at odd harmonics ($f_{\mathrm{in}}$, $3f_{\mathrm{in}}$, $5f_{\mathrm{in}}$, etc.); even harmonics are suppressed due to the symmetry of the capacitance-voltage relation \cite{Kozyrev2005}.  Focusing on the first and third harmonics, we define the respective peak values as $|v_k^f|$ and $|v_k^{3f}|$, and use these to plot Fig.~\ref{fig:profile}(a)--(b).  We verified that these experimental data agree well with results from the SPICE circuit simulator (see Supplemental Material \cite{SM}).

From Fig.~\ref{fig:profile}(a)--(b), we see that the nontrivial and trivial lattices exhibit very different behaviors for both the first- and third-harmonic signals.  First, consider the first-harmonic signal.  In both lattices, there is an exponential decay away from the edge, but the decay is sharper in the nontrivial lattice, which may be attributed to the enhanced intensity arising from the coupling of the input signal to the topological edge state.  As a quantitative measure of the localization of the first-harmonic signal, Fig.~\ref{fig:profile}(d)--(e) shows the inverse participation ratio (IPR) $\sum_k|v_k^f|^4 / (\sum_k |v_k^f|^2)^2$; a larger IPR corresponds to a more localized profile \cite{thouless1974}.  We see that the IPR is substantially larger in the nontrivial lattice than in the trivial lattice, over a broad range of $f_{\mathrm{in}}$ and $V_{\mathrm{in}}$.  The strong difference in localization is a key signature of nonlinearity: in the linear regime, a driving voltage on the edges of the nontrivial and trivial lattices would produce different overall amplitudes, but the same exponential decay profile \cite{SM}.  It is interesting to note that the region of enhanced IPR, shown in Fig.~\ref{fig:profile}(d), closely resembles the nontrivial bandgap in Fig.~\ref{fig:Circuit}(d).

\begin{figure}
  \includegraphics[width=0.49\textwidth]{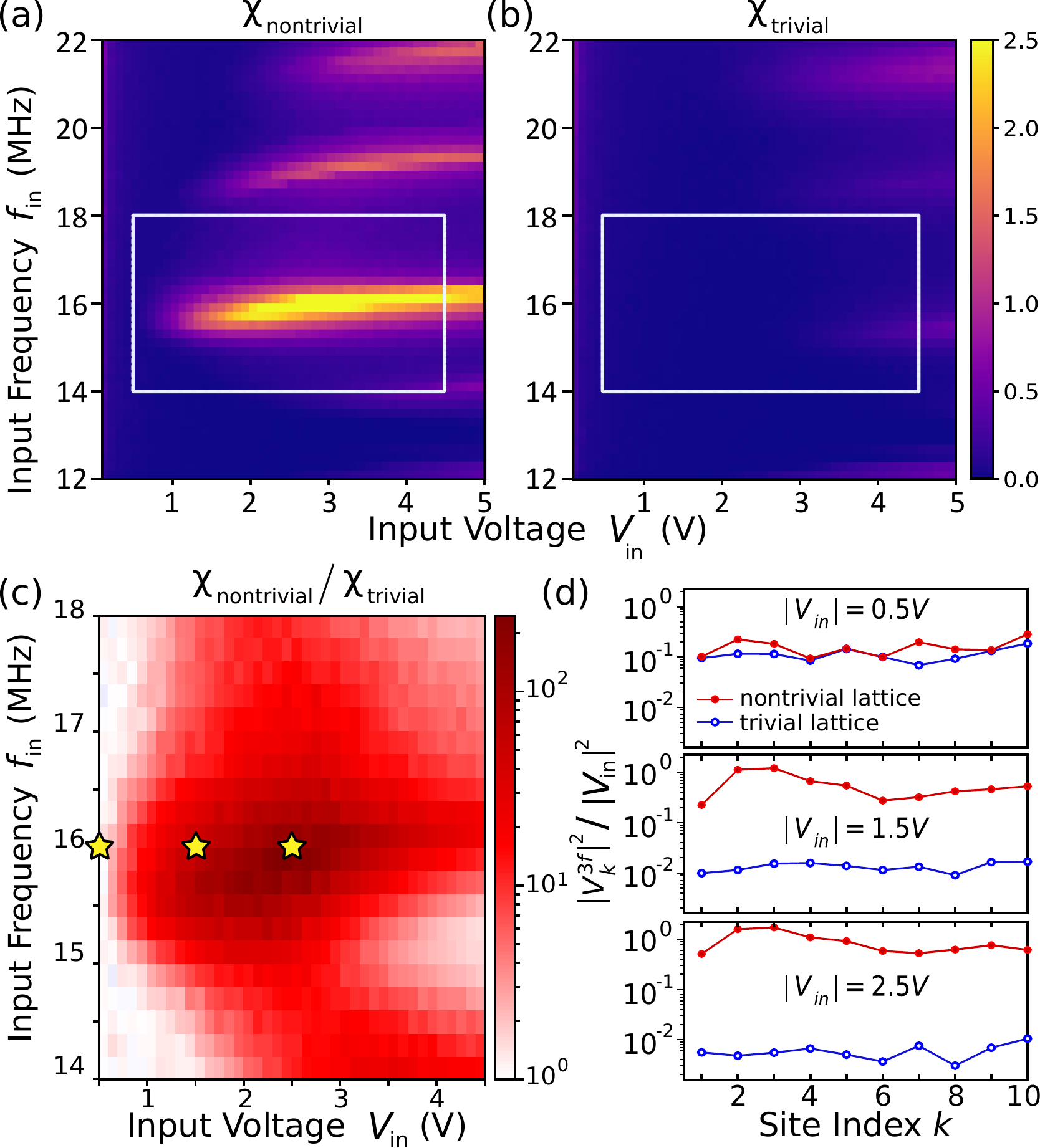}
  \caption{(a)--(b) Normalized third-harmonic signal intensity $\chi$ versus input frequency $f_{\mathrm{in}}$ and input voltage $V_{\mathrm{in}}$, for the (a) nontrivial and (b) trivial lattice.  Here, $\chi$ is derived from experimental data using the definition \eqref{eq:efficiency}.  (c) Ratio of the trivial and nontrivial intensities, $\chi_{\mathrm{nontrivial}}/\chi_{\mathrm{trivial}}$, within the region indicated by boxes in (a) and (b).  (d) Measured third-harmonic intensities (normalized to the input signal) at different sites, for the three sets of input parameters indicated by stars in (c): $V_{\mathrm{in}} = 0.5\,\mathrm{V}$, $1.5\,\mathrm{V}$, and $2.5\,\mathrm{V}$, with fixed $f_{\mathrm{in}}= 16\,\mathrm{MHz}$.}
  \label{fig:THG_eff}
\end{figure}

We can also see from Fig.~\ref{fig:profile}(a) and (c) that strong higher-harmonic signals are present in the nontrivial lattice.  Moreover, Fig.~\ref{fig:profile}(a) indicates that the third-harmonic signal is extended, not localized to the edge.  To understand this in more detail, we define
\begin{equation}
  \chi = \left\langle \left|v_k^{3f}\right|^2 \right\rangle / V_{\mathrm{in}}^2,
  \label{eq:efficiency}
\end{equation}
which quantifies the intensity of the third-harmonic signal relative to the input intensity at the first harmonic.  Here, $\langle\cdots\rangle$ denoting an average over the first ten lattice sites.  Fig.~\ref{fig:THG_eff}(a)--(b) plots the variation of $\chi$ with $f_{\mathrm{in}}$ and $V_{\mathrm{in}}$.  In the nontrivial circuit, the maximum value of the normalized intensity is $\chi \approx 2.5$ for $f_{\mathrm{in}} \sim 16\,\mathrm{MHz}$ and $1\,\mathrm{V} \lesssim V_{\mathrm{in}} \lesssim 4\,\mathrm{V}$.  The fact that $\chi$ peaks over a relatively narrow frequency range, as shown in Fig.~\ref{fig:THG_eff}(a), may be a finite-size effect: the high-frequency modes of the lattice form discrete sub-bands due to the finite lattice size [see Fig.~\ref{fig:Circuit}(d)--(e)].  In computer simulations, we obtained a similar maximum value of $\chi \approx 2.4$ for the nontrivial lattice, whereas a comparable left-handed NLTL of the usual design (containing only identical nonlinear capacitances) has maximum $\chi \approx 0.47$ \cite{SM}.

The trivial lattice exhibits a much weaker third-harmonic signal.  As indicated in Fig.~\ref{fig:THG_eff}(c), for certain choices of $f_{\mathrm{in}}$ and $V_{\mathrm{in}}$, the value of $\chi$ in the nontrivial lattice is $200$ times that in the trivial lattice.  Fig.~\ref{fig:THG_eff}(d) plots the normalized third-harmonic signal intensities versus the site index $k$, showing that they do not decay exponentially away from the edge.  In the nontrivial lattice, the normalized third-harmonic signal increases with $V_{\mathrm{in}}$ (i.e., stronger nonlinearity).

\textit{Discussion}---Our results point to a complex interplay between the topological edge state and higher-harmonic modes in the SSH-like NLTL.  When a topological edge state exists in the linear lattice, it can be excited by an input signal at frequencies matching the bandgap of the linear lattice. The importance of the edge state is evident from the comparisons between the topologically trivial and nontrivial lattices (Fig.~\ref{fig:profile} and Fig.~\ref{fig:THG_eff}).  Note also that when the excitation frequency lies outside the linear bandgap, the two lattices behave similarly and the harmonic generation is relatively weak.

In the topologically nontrivial lattice, the resonant excitation generates third- and higher-harmonic signals that penetrate deep into the lattice, unlike the first-harmonic mode which is localized to the edge.  Away from the edge, the higher-harmonic signals become stronger than the first harmonic, and hence dominate the effective value of the nonlinear $\alpha$ parameter.  In the linear lattice, $\alpha$ is the parameter that ``drives'' the topological transition, and increasing $\alpha$ leads to a larger bandgap and hence a more confined edge state.  In the nonlinear regime, Fig.~\ref{fig:profile} shows an order-of-magnitude increase in the third-harmonic signal amplitude in the nontrivial lattice, relative to the trivial lattice; this implies an effective increase in $\alpha$, and indeed we see that the first-harmonic mode profile is more strongly localized.  A more localized edge state, in turn, produces a stronger response to an input signal.

The above interpretation is supported by a more detailed analysis of the coupled equations governing the different circuit mode harmonics (see Supplemental Material \cite{SM}).  These equations involve an effective $\alpha$ parameter whose approximate value, in the $n$-th unit cell, is $\langle\alpha_n\rangle \approx A + 2B\sum_{m} |W_n^m|^2$, where $|W_n^m|$ is the $m$-th harmonic of the bias voltage on the nonlinear capacitor in the $n$-th unit cell, and $m = 1,3,5,\dots$  We are able to show that propagating waves can be self-consistently realized for higher ($m \ge 3$)  harmonics in the presence of non-linearity, even if the fundamental ($m=1$) mode only has decaying solutions.  The first-harmonic mode is localized to the edge, with localization length decreasing with $\langle\alpha_n\rangle$ in a manner similar to the linear SSH-like lattice.  The generation of the higher-harmonic signals occurs mainly near the edge of the lattice, where the first-harmonic mode is largest.  The nonlinearity-induced harmonic generation is aided by the well-known fact that the SSH edge state changes sign in each unit cell, corresponding to the fact that the gap closing in the SSH model takes place at the corner of the Brillouin zone \cite{ssh1979}.  This feature increases the bias voltages across the nonlinear capacitors, which can thus exceed the values of the voltages at individual sites.

The input signal can also be applied to the middle of the lattice.  In this context, it is interesting to note that when we choose to excite a single site in the bulk of an SSH-like lattice, the sections to either side of the excitation have different topological phases: either trivial on the left and nontrivial on the right, or vice versa, depending on the two possible choices of excitation site.  If the source impedance is sufficiently low, the effect is similar to exciting independent chains to the left and right; thus, the enhanced higher-harmonic signal is preferentially emitted toward the topologically nontrivial side (see Supplemental Material \cite{SM}).

The presence of higher-harmonic signals distinguishes our system from previous studies of nonlinear topological edge states, which were based on nonlinear \textit{self-modulation} at a single harmonic.  For instance, in a nonlinear SSH lattice where the coupling depends on the local intensity of a single mode, soliton-like edge states with anomalous mode profiles were predicted \cite{hadad2016}, and subsequently verified using a NLTL-like circuit \cite{hadad2018}.  That circuit, unlike ours, had narrow frequency bands and thus did not support propagating higher-harmonic modes.  Topological solitons based on nonlinear self-modulation are also predicted to exist in higher-dimensional lattices \cite{Lumer2013,Ablowitz2014, Leykam2016, Lumer2016, Kartashov2016, Gulevich2017}.  In our case, the effective value of $\alpha$ away from the edge is dominated by the higher-harmonic signals; from the point of view of the first-harmonic mode, these act as a \textit{nonlocal} nonlinearity, driving the entire lattice deeper into the topologically nontrivial regime, not just the sites with large first-harmonic intensity.

Our work opens the door to the application of topological edge states for enhancing harmonic generation, not just in transmission line circuits, but also a variety of other interesting systems.  These include two-dimensional electronic lattices, where topological edge states have already been observed in the linear regime \cite{Ningyuan2015}, and the unidirectional nature of the edge states may be even more beneficial for frequency-mixing \cite{Peano2016}. Higher dimensional circuit lattices may possess different thresholds for bulk propagation in different directions, with an extreme generalization being that of a corner mode circuit constructed in Ref.~\cite{lee2017}. Electronic circuits incorporating amplifiers and resistances may also be able to explore behaviors analogous to topological lasers \cite{R1Q1_1,R1Q1_2,R1Q1_3,R1Q1_4,R1Q1_5}, combining topological states with both nonlinearity and non-Hermiticity. Finally, circuits containing varactors that are explicitly time-modulated may be suitable for generating synthetic dimensions to realize topological features in higher dimensions \cite{R1Q2_1,R1Q2_2,R1Q2_3,R1Q2_5,R1Q2_6,R1Q2_7}.

\textit{Methods}---The nonlinear transmission line was implemented on a PCB (Seeed Tech.~Co), with each nonlinear capacitor consisting of a pair of back-to-back varactors (Skyworks Solutions, SMV1253-004LF).  The transmission line, as fabricated, is topologically nontrivial, as shown in Fig.~\ref{fig:Circuit}(a).  To probe the trivial circuit, we use a switch to add one sublattice unit cell at the rightmost end of the transmission line, and disconnect the leftmost $C_a$ and $L$ in Fig.~\ref{fig:Circuit}(a).  This yields a nontrivial circuit of same length, with the $C_a$ and $C_b$ capacitors swapped.

A function generator (Tektronix AFG3102C) supplies the continuous-wave sinusoidal input voltage, and the voltages on successive lattice sites, $k \ge 1$, are measured by an oscilloscope (Rohde \& Schwarz RTE1024) in high-impedance mode.  Numerical results were obtained using the SPICE circuit simulator.

\textit{Acknowledgments}---We are grateful to H.~Wang and D.~Leykam for helpful discussions.  YW, LJL, BZ, and CYD were supported by the Singapore MOE Academic Research Fund Tier 2 Grant MOE2015-T2-2-008, and the Singapore MOE Academic Research Fund Tier 3 Grant MOE2016-T3-1-006.

\begin{widetext}

\pagebreak

  \begin{center}
{\large Supplemental Material for: \\ Topologically Enhanced Harmonic Generation in a Nonlinear Transmission Line Metamaterial}    
  \end{center}

\setcounter{figure}{0}
\renewcommand{\thesection}{S\arabic{section}}
\renewcommand{\thefigure}{S\arabic{figure}}

\section{Linear circuit}

Consider the circuit shown in Fig.~\ref{fig:Circuit2}(a) below.  This is the same as Fig.~1(a) of the main text, but with the complex voltage variables relabeled for convenience. The band diagram, plotted in Fig.~\ref{fig:Circuit2}(b), shows that the upper band exhibits negative dispersion, regardless of the value of $\alpha$. On unit cell $n$, the voltages on the two sites are $v_{n}^a$ (to the right of the $C_a$ capacitor) and  $v_{n}^b$ (to the right of the $C_b$ capacitor).  Also, we let $i^a_n$ ($i^b_n$) denote the current through the inductor to the right of the $C_a$ ($C_b$) capacitor.

\begin{figure}[h]
  \includegraphics[width=0.95\textwidth]{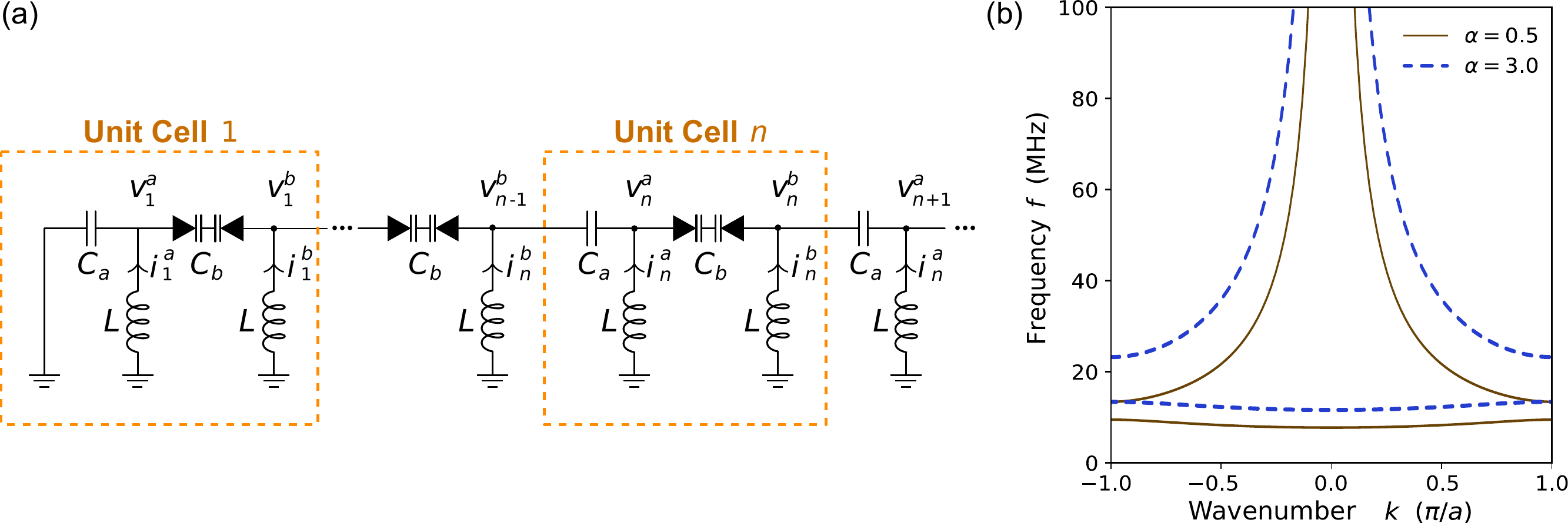}
  \caption{ (a) Schematic of the left-handed transmission line circuit. (b) Band diagram for the infinite lattice.  Results are shown for $\alpha=0.5$ (solid curves) and $3.0$ (dashes).}
  \label{fig:Circuit2}
\end{figure}

Let $C_a$, $C_b$, and $L$ be constants, and take a harmonic mode with angular frequency $\omega$.  For $n > 1$, we apply Kirchhoff's laws to the inductors and capacitors, with the $\exp(i\omega t)$ phasor convention, and obtain
\begin{align}
  i \omega L i_n^a + v_n^a &= 0 \label{eq:K1}\\
  i \omega L i_n^b + v_n^b &= 0 \\
  -i\omega \, C_a\left(v_n^a-v_{n-1}^b\right)
  + i_n^a + i\omega \, C_b \left(v_n^b-v_n^a\right) &= 0 \\
  -i\omega \, C_b\left(v_n^b-v_{n}^a\right)
  + i_n^b + i\omega \, C_a \left(v_{n+1}^a-v_n^b\right) &= 0.
  \label{eq:K4}
\end{align}
Combining these to eliminate $i^a_n$ and $i^b_n$ yields the following pair of coupled equations:
\begin{align}
  v_{n-1}^b - \frac{1}{\alpha} v_n^a + \frac{1}{\alpha} v_n^b &=
  \left(1-\Omega^2\right) v_n^a \\
  \frac{1}{\alpha} v_{n}^a - \frac{1}{\alpha} v_n^b + v_{n+1}^a &= 
  \left(1-\Omega^2\right) v_n^b,
\end{align}
where $\Omega^2 \equiv \omega_a^2/\omega^2$, $\omega_a \equiv 1/\sqrt{LC_a}$ and $\alpha \equiv C_a/C_b$.

Suppose we close the circuit by grounding the leftmost and rightmost sites.  Consider the left edge (the right edge is handled similarly).  There, the Kirchhoff equations simplify to
\begin{align}
  i \omega L i_1^a + v_n^a &= 0 \\
  -i\omega \, C_a \, v_1^a
  + i_1^a + i\omega \, C_b \left(v_1^b-v_1^a\right) &= 0,
\end{align}
resulting in the boundary equation
\begin{equation}
  - \frac{1}{\alpha} v_1^a + \frac{1}{\alpha} v_1^b = 
  \Big(1-\Omega^2\Big) v_1^a.
  \label{boundary}
\end{equation}
Hence, we arrive at the modified SSH problem discussed in the main text:
\begin{equation}
  \begin{pmatrix}
    -\frac{1}{\alpha}&\frac{1}{\alpha}&&&&\\
    \frac{1}{\alpha}&-\frac{1}{\alpha}&1&&&\\
    &1&-\frac{1}{\alpha}&\frac{1}{\alpha}&&\\
    &&\frac{1}{\alpha}&-\frac{1}{\alpha}&\ddots&\\
    &&&\ddots&\ddots&\\
  \end{pmatrix}
  \begin{pmatrix}
    v_1^a\\v_1^b\\v_2^a\\ v_2^b \\ \vdots
  \end{pmatrix}
  = \Big(1 -\Omega^2\Big)
  \begin{pmatrix}
    v_1^a\\v_1^b\\v_2^a\\v_2^b \\\vdots
  \end{pmatrix}.
  \label{eigvaleq}
\end{equation}
This is the configuration referred to in the main text as the ``nontrivial lattice''.  The ``trivial lattice'' can be described by removing the first row and column of the matrix.  In either case, the edges of the band gap are
\begin{equation}
  \omega_\pm = \left\{\sqrt{\frac{\alpha}{2}}\omega_a \;\;\; \mathrm{or} \;\;\;
  \frac{\omega_a}{\sqrt{2}}\right\},
\end{equation}
and the angular frequency of the edge state is
\begin{equation}
  \omega_{\mathrm{es}} = \sqrt{\alpha/(1+\alpha)}\, \omega_a.
  \label{edgefreq}
\end{equation}

\begin{figure}[b]
  \includegraphics[width=0.85\textwidth]{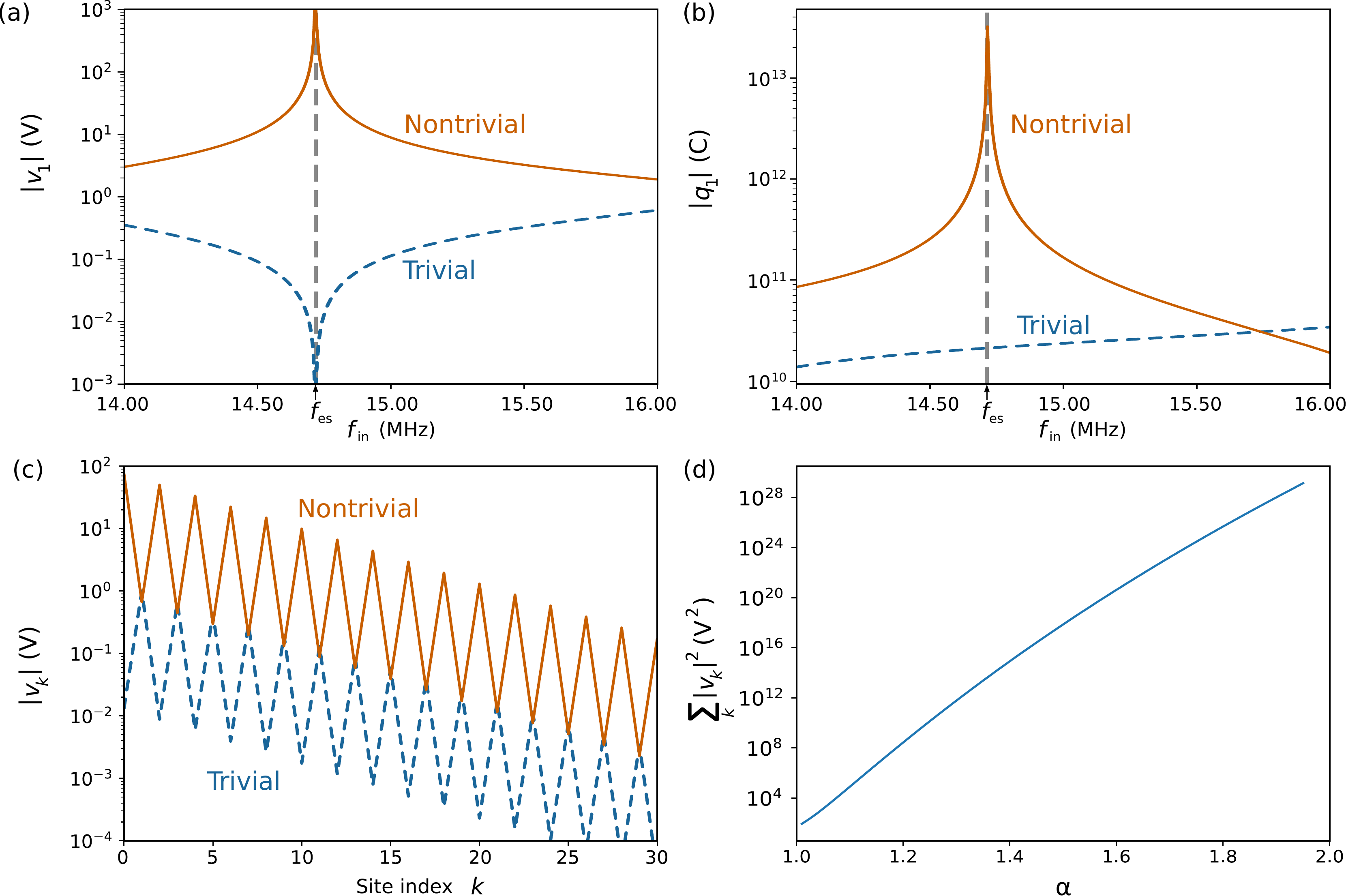}
  \caption{Response of the linear circuit to an external voltage, calculated numerically from the circuit equations with the parameters $f_a = 19\,\mathrm{MHz}$ and $C_a = 47\,\mathrm{pF}$.  (a)--(b) Resonant response of (a) the amplitude of the voltage on the first site (to the right of the leftmost capacitor), and (b) the amplitude of the stored charge on the first (leftmost) capacitor, versus source frequency $f_{\mathrm{in}} = \omega/2\pi$, for $\alpha = 1.5$.  The resonance frequency predicted by Eq.~\eqref{edgefreq} is indicated by the horizontal dashed line.  (c) Spatial distribution of the voltage ampltidues for $f_{\mathrm{in}} = 14.75\,\mathrm{MHz}$ and $\alpha = 1.5$. (d) Intensity of the nontrivial lattice's resonant mode, as given by the value of $\sum_k|v_k|^2$ (where $|v_k|$ is the voltage amplitude at site $k$) at resonance, versus the $\alpha$ parameter.  For each value of $\alpha$, the input voltage has amplitude $1\,\mathrm{V}$ and frequency given by Eq.~\eqref{edgefreq}. }
  \label{fig:linear}
\end{figure}

Next, we consider the response of the circuit to a harmonic voltage source.  Instead of grounding the left edge, we apply an input voltage of amplitude $V_{\mathrm{in}}$ and frequency $\omega$.  Then Eq.~\eqref{boundary} is replaced by
\begin{equation}
  V_{\mathrm{in}} - \frac{1}{\alpha} v_1^a + \frac{1}{\alpha} v_1^b = 
  \Big(1-\Omega^2\Big) v_1^a,
\end{equation}
and the eigenvalue equation \eqref{eigvaleq} is replaced by an inhomogenous equation.

Numerical solutions for this problem are shown in Fig.~\ref{fig:linear}.  Fig.~\ref{fig:linear}(a)--(b) shows that the voltage amplitude in the nontrivial lattice is resonantly enhanced when $\omega$ matches $\omega_{\mathrm{es}}$.  The trivial lattice, on the other hand, does not exhibit a resonant enhancement.  However, when we plot the voltage distributions, they have the same decay constant, as shown in Fig.~\ref{fig:linear}(c).  This is due to the fact that they have the same \textit{bulk} Hamiltonians.

Fig.~\ref{fig:linear}(d) shows the modal intensity (the sum of squared voltage amplitudes over the lattice) at resonance, versus the $\alpha$ parameter, for a nontrivial lattice.  In this plot, the input frequency for each value of $\alpha$ is adjusted to the edge state frequency given by Eq.~\eqref{edgefreq}.  With increasing $\alpha$, the bandgap becomes larger (i.e., the lattice moves deeper into the topologically nontrivial phase); accordingly, the edge state is more strongly confined, and responds more strongly to the resonant excitation.  The modal intensity scales exponentially with the bandgap size.

\section{Nonlinear circuit}

\subsection{Circuit equations}
\label{sec:nonlin_circuit}

We seek a set of time-domain equations for the circuit's nonlinear regime, where the $B$ capacitors are nonlinear.  Let $q_n^a(t)$ and $q_n^b(t)$ denote the charges stored in capacitors $A$ and $B$, respectively, on site $n$.  These obey
\begin{align}
  q_n^a(t) &= C_a \Big[v_n^a(t) - v_{n-1}^b(t)\Big], \label{qvars1}\\
  q_n^b(t) &= C_{bn}(t) \Big[v_n^b(t) - v_{n}^a(t)\Big]. \label{qvars2}
\end{align}
Here, $C_{bn}(t)$ is the value of the nonlinear $B$ capacitance in unit cell $n$.  Using Kirchhoff's laws, we can derive several additional equations.  The time-dependent voltage-current relations on the inductors are
\begin{align}
  v_n^a &= - L \frac{di_n^a}{dt}, \label{inductors1} \\
  v_n^b &= - L \frac{di_n^b}{dt}. \label{inductors2}
\end{align}
The current-charge relations on the capacitors, with the assumptions of current conservation and zero net charge, give
\begin{align}
  \frac{dq_n^a}{dt} - \frac{dq_n^b}{dt} &= i_n^a \label{dqdt1} \\
  \frac{dq_n^b}{dt} - \frac{dq_{n+1}^a}{dt} &= i_n^b. \label{dqdt2}
\end{align}
By combining Eqs.~\eqref{qvars1}--\eqref{dqdt2}, we can eliminate the $q^{a/b}$ and $i^{a/b}$ variables, resulting in the following pair of time-domain circuit equations expressed in terms of the $v^{a/b}$ variables:
\begin{align}
  -\frac{d^2}{dt^2}\left[v_n^a - v_{n-1}^b
    - \frac{1}{\alpha_n} \left(v_n^b - v_n^a\right)\right]
  &= \omega_a^2\, v_n^a(t) \label{nonlin1} \\
  -\frac{d^2}{dt^2}\left[v_n^b -v_{n+1}^a
    + \frac{1}{\alpha_n} \left(v_{n}^b - v_{n}^a\right)
    \right]
  &= \omega_a^2 \, v_n^b(t). \label{nonlin2}
\end{align} 
Here,
\begin{equation}
  \alpha_n(t) \equiv \frac{C_a}{C_{bn}(t)} \label{alphan}
\end{equation}
is the nonlinear capacitance ratio at site $n$.

It is convenient to re-cast Eqs.~\eqref{nonlin1}--\eqref{nonlin2} in terms of the variables
\begin{align}
  u_n &= v_n^b + v_n^a \\
  w_n &= v_n^b - v_n^a.
\end{align}
Then
\begin{align}
  -\frac{d^2}{dt^2} \left[-\frac{1}{2}u_{n+1} + u_n - \frac{1}{2}u_{n-1}
    + \frac{w_{n+1} - w_{n-1}}{2} \right] &= \omega_a^2 u_n \label{nonlinu} \\
  -\frac{d^2}{dt^2} \left[-\frac{u_{n+1} - u_{n-1}}{2}
    + \frac{w_{n+1} + w_{n-1}}{2} + \left(1 + \frac{2}{\alpha_n(t)}\right) \, w_n
    \right] &= \omega_a^2 w_n.
  \label{nonlinw}
\end{align}

\subsection{Harmonic decomposition and nonlinearity model}
\label{sec:hdecomp}

If a harmonic signal is injected into the nonlinear circuit, higher harmonics are generated.  Due to the symmetric C-V curve of the nonlinear capacitors, even-order harmonics are suppressed.

Let $\omega$ denote the frequency of the first harmonic.  We will decompose the voltage variables in the following way:
\begin{align}
  u_n(t) &\approx \sum_{m = 1, 3, 5, \cdots} (-1)^n \;
  U_{n}^{m} \, e^{im\omega t} \;\;\,+\; \mathrm{c.c.} \label{uwansatz1}\\
  w_n(t) &\approx \sum_{m = 1, 3, 5, \cdots} (-1)^n \;
  W^m_{n} \, e^{im\omega t} \;+\; \mathrm{c.c.}
  \label{uwansatz2}
\end{align}
On the right hand sides, the integer superscripts $\{1, 3, 5, \dots\}$ denote the harmonic index.  The factor of $(-1)^n$ is for later convenience; we expect the first harmonic mode to behave like an SSH edge state, which is characterized by alternating signs on adjacent unit cells (another way of saying this is that the band gap of the bulk SSH model is narrowest at the Brillouin zone boundary, $k = \pm\pi/h$, where $h$ is the lattice constant), and this factor ensures that the $U_n^1$ and $W_n^1$ variables act as smooth envelopes with the sign alternation taken out.

We now have to substitute the ansatz \eqref{uwansatz1}--\eqref{uwansatz2} into Eqs.~\eqref{nonlinu}--\eqref{nonlinw}.  First, consider Eq.~\eqref{nonlinu}, which is easy to deal with since it is linear.  Matching the individual harmonics, we obtain
\begin{align}
  \frac{1}{2} U_{n+1}^1 + U_n^1 + \frac{1}{2} U_{n-1}^1
    - \frac{1}{2}\left(W_{n+1}^1 - W_{n-1}^1\right) &= \Omega_1^2 \;U_n^1,
  \label{linearpart1}
\end{align}
where
\begin{equation}
  \Omega_m^2 \equiv \frac{\omega_a^2}{m^2\omega^2}.
\end{equation}

Next, consider the nonlinear equation \eqref{nonlinw}.  The main complication here is the term involving
\begin{equation}
  \frac{w_n(t)}{\alpha_n(t)}.
  \label{wovera}
\end{equation}
The time variation of $\alpha_n(t)$ gives rise to two classes of effects: (i) self-phase modulation and cross-phase modulation, which alter the effective value of $\alpha_n$ ``seen'' by each given harmonic, and (ii) frequency-mixing processes, which couple the dynamical equations for the different harmonics.  For now, let us try to pick out the contributions to category (i), neglecting (ii).

In our experiment, each nonlinear capacitor consists of a pair of back-to-back varactors.  The nonlinear capacitance ratio $\alpha_n$ was defined in Eq.~\eqref{alphan}.  Let us make the assumption that
\begin{equation}
  \alpha_n(t) \;\approx\; A + B \left[v_n^b(t) - v_n^a(t)\right]^2
  \;=\; A + B \left[w_n(t)\right]^2.
  \label{nonlinearity_model}
\end{equation}
Here, $A$ is the capacitance ratio in the linear limit and $B > 0$ is a Kerr-like parameter determining the strength of the lowest-order nonvanishing (cubic) nonlinearity.  To obtain values for $A$ and $B$, we use the manufacturer-supplied capacitance-voltage curve for the individual varactors to calculate $\alpha$ and the bias voltage $\Delta V$ for a pair of back-to-back varactors.  We then perform a linear least-squares fit of $\alpha$ versus $\Delta V^2$, using the subset of data points with voltage biases $\Delta V \le 1\,\mathrm{V}$.  The fitted parameters are $A = 1.32$ and $B = 0.51\,\mathrm{V}^{-2}$, and the fit is shown in Fig.~1(c) of the main text.

In \eqref{wovera}, we can take the approximation of replacing $\alpha_n(t)$ with its time-independent part,
\begin{equation}
  \langle\alpha_n\rangle \approx
  A + 2B\!\! \sum_{m = 1, 3, \dots} \Big| W_{n}^m \Big|^2.
  \label{aavg}
\end{equation}
For each harmonic $m$, this would then give rise to a term
\begin{equation}
  \propto \frac{W_n^m}{\langle\alpha_n\rangle} \; e^{im\omega t},
  \label{effectiveterm}
\end{equation}
with $\langle\alpha_n\rangle$ now playing the role of an ''effective'' $\alpha$ parameter.

This approximation does not capture all possible self-phase and cross-phase modulation terms.  This can be seen in the low-intensity limit, where we can Taylor expand $1/\alpha_n(t)$ in the $W_n^m$ variables; in this expansion, there will be non-constant terms like $W_n^m (W_n^{m'})^* e^{i(m-m')\omega t}$, which couples to the $W_n^{m'} e^{im'\omega t}$ harmonic term from $w_n(t)$ to yield a term proportional to $W_n^m\, e^{im\omega t}$, and hence contributing to the self-phase or cross-phase modulation.  We will not undertake a rigorous analysis of these terms, since the Taylor expansion is invalid anyway when the intensities are not small.  Instead, our take-home message is as follows:

\begin{enumerate}
\item Each harmonic contributes to the effective value of $\alpha$ in direct proportion to its local intensity, like in Eq.~\eqref{aavg}.

\item However, the precise numerical factor need not be exactly the same as in Eq.~\eqref{aavg}.
\end{enumerate}

Based on this approximation, we can now deal with the nonlinear equation \eqref{nonlinw}, which simplifies to
\begin{equation}
  \frac{U^1_{n+1} - U^1_{n-1}}{2}
  - \frac{W^1_{n+1} + W^1_{n-1}}{2} + \left(1 +
    \frac{2}{\langle\alpha_n\rangle}
    \right) \, W_n^1 = \Omega_1^2 \; W^1_n.
    \label{nonlinearpart1}
\end{equation}

\subsection{Localized and traveling-wave solutions}

Let us consider the case where $\langle \alpha_n\rangle$ is approximately constant in space, and look for solutions of the form
\begin{equation}
  U_n^m = U_m e^{ik_m n}, \quad W_n^m = W_m e^{ik_m n}.
\end{equation}
These are traveling-wave solutions if $k_m$ is real, and exponentially localized solutions if $k_m$ is complex.  Substituting this into Eqs.~\eqref{linearpart1} and \eqref{nonlinearpart1} gives
\begin{equation}
  \begin{pmatrix} 1 + \cos k_m & -i\sin k_m \\
    i\sin k_m & 1 + \frac{2}{\langle\alpha\rangle} - \cos k_m
  \end{pmatrix}
  \begin{pmatrix} U_m \\ W_m \end{pmatrix}
  = \Omega_m^2 \begin{pmatrix} U_m \\ W_m \end{pmatrix}.
\end{equation}
Solving the characteristic equation gives
\begin{equation}
  \cos k_m = \Big(1+\langle\alpha\rangle\Big)\Omega_m^2 
  - \Big(1 + \frac{\langle\alpha\rangle}{2}\Omega_m^4\Big).
\end{equation}
We can then easily show that, for $\langle \alpha\rangle > 1$, the domains over which the right-hand side has magnitude smaller than unity (i.e., $k_m$ is real) are:
\begin{align}
  \Omega_m^2 &< 2/ \langle \alpha\rangle \label{ineq1}\\
  2 &< \Omega_m^2 < 2(1+1/\langle \alpha\rangle).
\end{align}
For $m = 1$, this corresponds exactly to the bands shown in Fig.~1(d)--(e) of the main text.  In particular, within the band gap between $\omega_a/\sqrt{2}$ and $\sqrt{\alpha/2}\,\omega_a$, the right-hand side is larger than unity and hence $k_1$ is imaginary, in complete agreement with the linear analysis.

For the higher-harmonic modes, \eqref{ineq1} is satisfied easily.  For example, for the third harmonic, we require
\begin{equation}
  \omega^2 > \frac{\langle \alpha\rangle}{18} \omega_a^2.
\end{equation}
For operating frequencies below the linear-regime band gap, $\omega < \sqrt{A/2} \;\omega_a$, this is satisfied for
\begin{equation}
  \alpha < 9 A,
\end{equation}
which is well within the regime considered in this experiment.  This analysis thus confirms that the nonlinear circuit is capable of supporting traveling-wave higher-harmonic solutions.

\subsection{SPICE simulation results: voltage profiles}
\label{sec:spice1}

We performed simulations of the nonlinear circuit using the circuit simulation software SPICE.  The simulations reproduce the basic features of the experimental results, though the results are not in exact agreement, probably due to imperfections in the circuit components.

\begin{figure*}
  \includegraphics[width=0.65\textwidth]{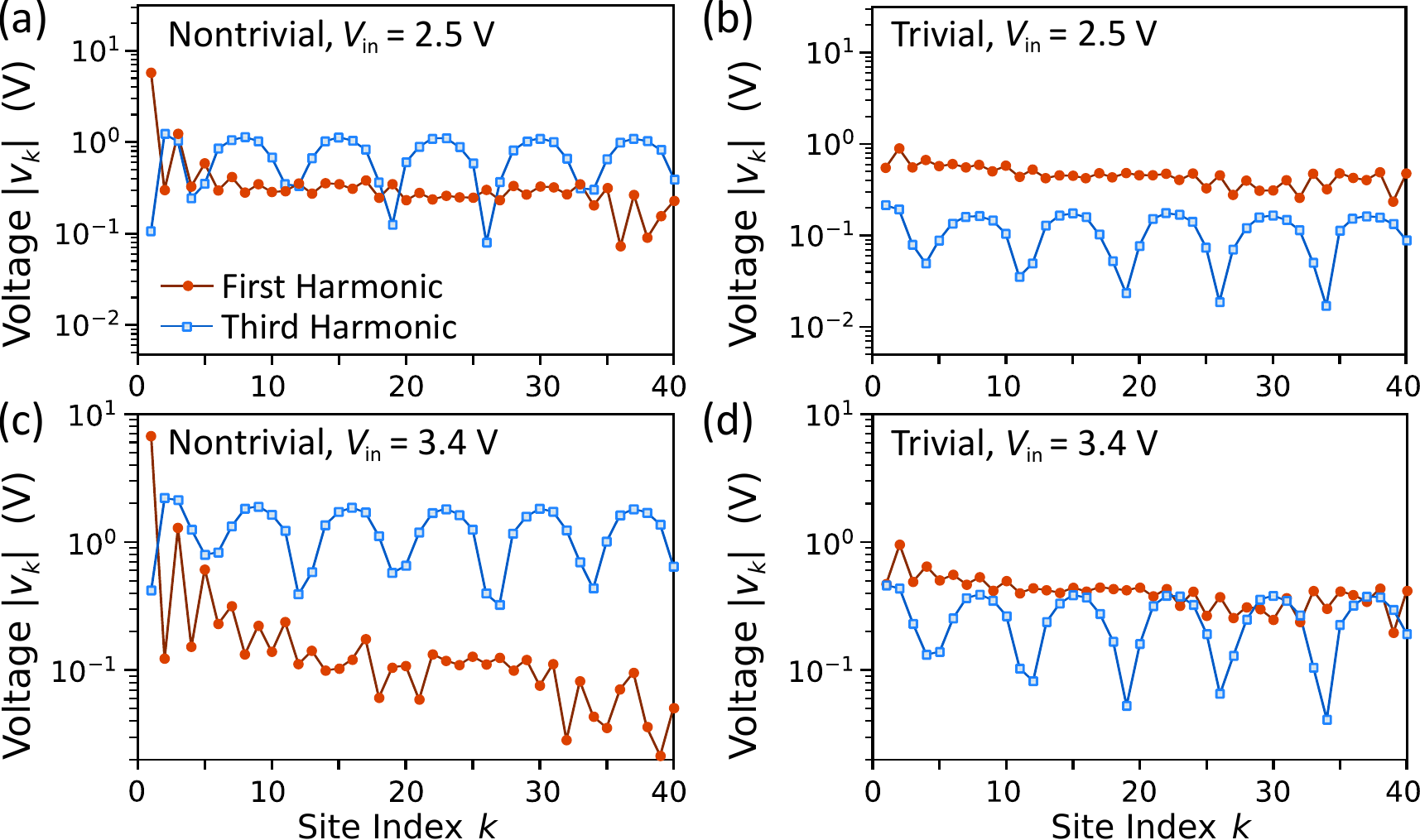}
  \caption{On-site voltage amplitudes for the first-harmonic signal (orange circles) and third-harmonic signal (blue squares).  The frequency of the input is $f_{\mathrm{in}} = 16.4\,\mathrm{MHz}$.  (a)--(b) Nontrivial and trivial lattices with $V_{\mathrm{in}} = 2.5\,\mathrm{V}$.  (c)--(d) Nontrivial and trivial lattices with $V_{\mathrm{in}} = 3.4\,\mathrm{V}$. }
  \label{fig:spice}
\end{figure*}

Fig.~\ref{fig:spice} shows the on-site voltage amplitudes for the first- and third-harmonic signals.  These are extracted from the simulation results in a manner similar to the experiment: after the simulation reaches steady-state, we take a time-dependent sample, Fourier transform, and extract the peak heights.  To obtain simulation results matching the experimental results shown in Fig.~2(a)--(b) of the main text, we find that it is necessary to apply a higher input voltage amplitude than in the experiment, $V_{\mathrm{in}} \approx 3.4\,\mathrm{V}$.  The results are shown in Fig.~\ref{fig:spice}(c)--(d).  Similar to the experiment, the first-harmonic mode in the nontrivial lattice decays away from the edge, reaching values much lower than in the trivial lattice.

\begin{figure*}
  \includegraphics[width=0.65\textwidth]{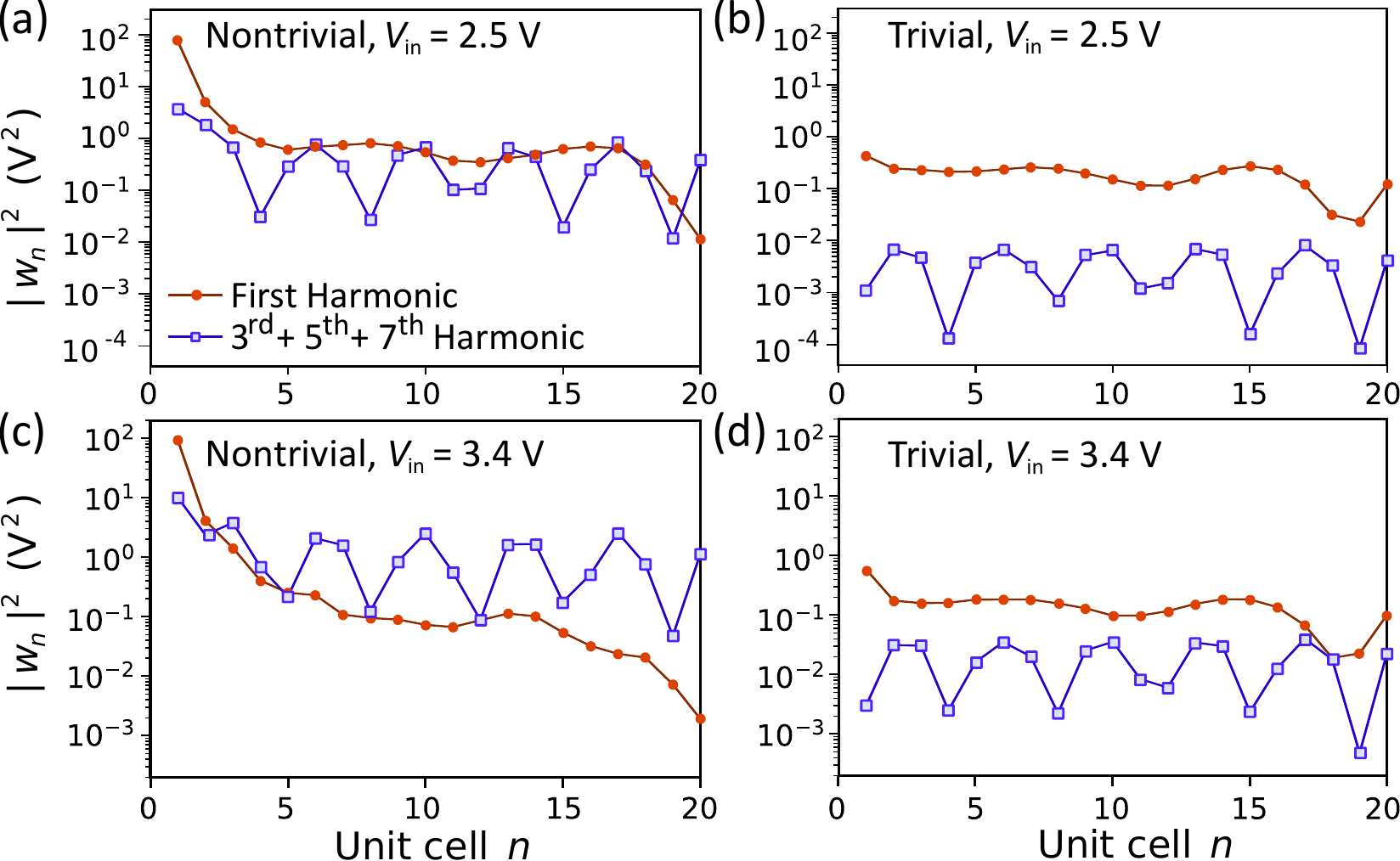}
  \caption{Squared bias voltage amplitudes on the nonlinear capacitors, for the first-harmonic signal (orange circles) and higher-harmonic signals (purple squares).  The frequency of the input is $f_{\mathrm{in}} = 16.4\,\mathrm{MHz}$.  The higher-harmonic data is obtained by summing over the third, fifth, and seventh harmonic data (further harmonics are negligible).  (a)--(b) Nontrivial and trivial lattices with $V_{\mathrm{in}} = 2.5\,\mathrm{V}$.  (c)--(d) Nontrivial and trivial lattices with $V_{\mathrm{in}} = 3.4\,\mathrm{V}$. }
  \label{fig:spice2}
\end{figure*}

Fig.~\ref{fig:spice2} shows the bias voltage amplitudes on the nonlinear capacitors, which were not measured in the experiment.  As discussed in Section~\ref{sec:hdecomp}, the bias voltages determine the effective value of the nonlinear $\alpha$ parameter.  To obtain this data from the simulations, we extract the time-dependent bias voltage samples (i.e., the time-dependent voltages between the ports of the nonlinear capacitors, denoted by $w_n(t)$ in Section~\ref{sec:nonlin_circuit}), Fourier transform, and extract the peak heights; this yields the components denoted by $|W_n^m|$ in Section~\ref{sec:hdecomp}.  According to Eq.~\eqref{aavg}, the contribution of each harmonic to the effective local $\alpha$ is proportional to $|W_n^m|^2$.  Fig.~\ref{fig:spice2} shows a comparison between the first-harmonic contribution (orange circles) and the higher-harmonic contributions (purple squares).  In particular, in Fig.~\ref{fig:spice2}(c), which correspond to the voltage plot of Fig.~\ref{fig:spice}(c), the higher-harmonic signals are found to increasingly dominate the nonlinearity as we go deeper into the lattice.

\subsection{SPICE simulation results: mid-lattice excitation}

If we choose to excite a site in the middle of the circuit, rather than the edge, the behavior depends on the input impedance of the voltage source.  First, consider a low-impedance voltage source.  In this case, the voltage on the input site is rigidly determined by the voltage source, so this is similar to exciting two independent transmission lines.

If lattice is uniform (i.e., defect free), once we pick an excitation site, the lattice sections to the left and right of the excitation site necessarily have different topological phases.  For instance, for $\alpha > 1$, if there is a $C_a$ capacitor to the right of the excitation site, then the section on the right is nontrivial and the section on the left is trivial.  Under such circumstances, reasoning from the behavior of the circuit under edge excitation, we expect the higher-harmonic signal to be emitted asymmetrically: a strong higher-harmonic signal should propagate to the topologically nontrivial side, with a weak higher-harmonic signal on the trivial side.  This prediction is verified by the SPICE simulation results plotted in Fig.~\ref{fig:middle}.

\begin{figure*}[h]
  \includegraphics[width=0.8\textwidth]{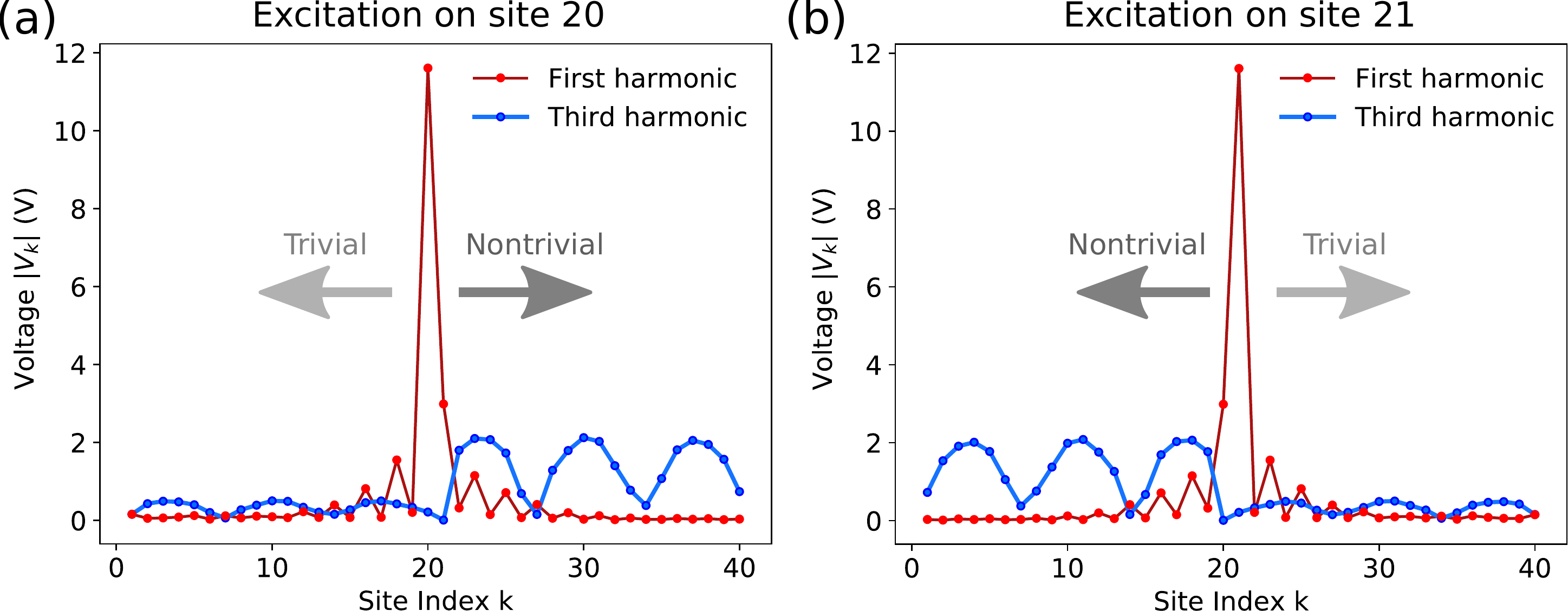}
  \caption{Voltage amplitude profiles for the first and third harmonic signals, from SPICE simulation on a 40-site NLTL with (a) site 20 excited, and (b) site 21 excited.  The choice of excitation site partitions the lattice into topologically trivial and nontrivial sections.  The input signal has voltage amplitude $V_{\mathrm{in}} = 3\,\mathrm{V}$, frequency $f_{\mathrm{in}} = 16\,\mathrm{MHz}$, and input impedance $1\Omega$.  All NLTL simulation parameters are the same as in Section~\ref{sec:spice1}.}
  \label{fig:middle}
\end{figure*}

When the input impedance of the voltage source is high, the behavior is less clear-cut, as both the first- and higher-harmonic signals can easily cross the excitation site.  In both the topologically trivial and nontrivial lattices, the higher-harmonic modes are in-band; thus, any higher-harmonic signal that is generated can propagate to either the trivial or nontrivial side.  At each frequency, the dominant direction of higher-harmonic emission will depend on the availability of circuit modes, based on finite-size effects, in each lattice section.

\subsection{SPICE simulation results: comparison with conventional NLTL}
 
Finally, we used SPICE simulations to compare the third harmonic intensity in this circuit to a conventional left-handed NLTL.  In the conventional left-handed NLTL, all the linear $C_a$ capacitors are replaced with nonlinear $C_b$ capacitors (i.e., the lattice is no longer dimerized). In the linear limit, the conventional left-handed NLTL with $C_b=35\,\textrm{pF}$ is a high-pass filter with a Bragg cutoff frequency of $11\,\textrm{MHz}$.

\begin{figure*}[h]
  \includegraphics[width=0.55\textwidth]{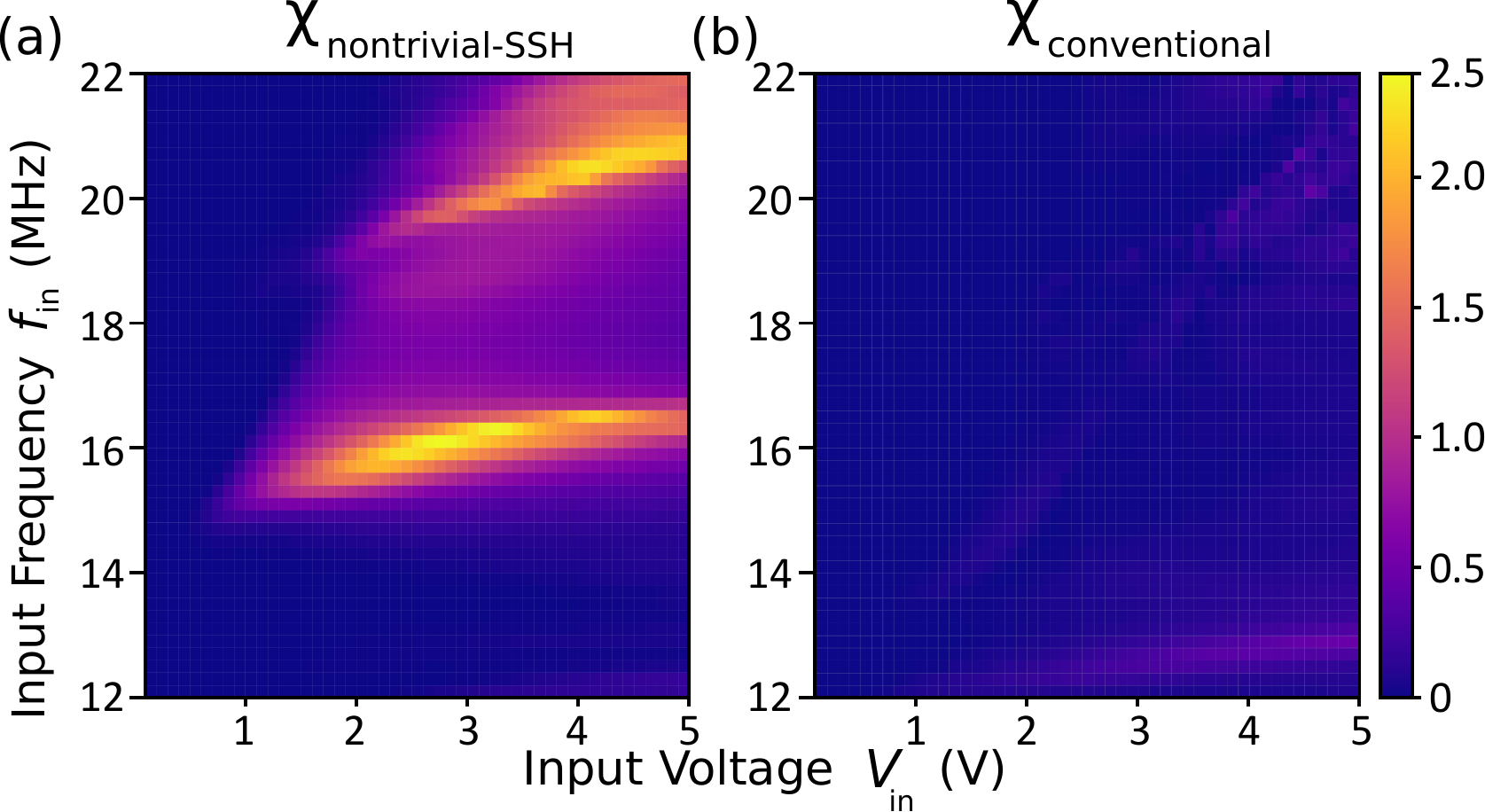}
  \caption{Plot of the normalized third-harmonic intensity $\chi$ versus input frequency $f_{\mathrm{in}}$ and input voltage $V_{\mathrm{in}}$, for (a) the SSH-like lattice in its topologically nontrivial configuration, and (b) a conventional left-handed nonlinear transmission line (NLTL) with identical nonlinear capacitors.  The figure of merit $\chi$ is defined in the same way as in the main text, as the mean squared third-harmonic amplitude on the first 10 sites relative to the squared input voltage amplitude.  In the SSH case, we obtain $\chi \gtrsim 2.4$, which is comparable to the experimental result of $\chi \approx 2.5$ reported in the main text.  In the conventional NLTL, we observe $\chi < 0.47$ throughout the parameter range.}
	\label{fig:THG_conventional}
\end{figure*}

Fig.~\ref{fig:THG_conventional} plots the simulation results for the normalized third-harmonic intensity $\chi$ (defined in the same way as in the main text), versus the input parameters $f_{\mathrm{in}}$ and input voltage $V_{\mathrm{in}}$.  The simulation results for the SSH-like lattice, shown in Fig.~\ref{fig:THG_conventional}(a), are similar to the experimental results shown in Fig.~3(a) of the main text; in particular, the maximum value of $\chi$ is $\gtrsim 2.4$, comparable to the experimentally-obtained maximum value $\chi \approx 2.5$.  By contrast, Fig.~\ref{fig:THG_conventional}(b) shows that the conventional NLTL exhibits no comparable enhancement of the third-harmonic signal intensity, with $\chi < 0.47$ throughout the entire parameter regime we investigated.  Hence, the introduction of the topological edge mode has contributed to a five-fold increase in the intensity of the generated third-harmonic signal.

\end{widetext}

\end{document}